\documentclass{PoS}

\title{Investigation of Ultra-Luminous Infrared Galaxies as Obscured High-Energy Neutrino Source Candidates}

\ShortTitle{Investigation of ULIRGs as Obscured Neutrino Sources}

\author{
The IceCube Collaboration\footnote{For collaboration list, see PoS(ICRC2019) 1177.}\\
{\itshape \href{http://icecube.wisc.edu/collaboration/authors/icrc19_icecube}{http://icecube.wisc.edu/collaboration/authors/icrc19\_icecube}}\\
E-mail: \email{pablo.correa@icecube.wisc.edu}
}

\abstract{

Ultra-Luminous Infrared Galaxies (ULIRGs) are the most luminous objects in the infrared sky. With infrared luminosities exceeding $10^{12}$ solar luminosities, ULIRGs contain strong star formation regions which could power hadronic acceleration. Moreover, a significant fraction of ULIRGs have been found to host Active Galactic Nuclei, which could also be a source of hadronic acceleration. Furthermore, such high infrared luminosities indicate that large amounts of dust are present in these objects. In the presence of hadronic acceleration, this dust not only represents an excellent target for high-energy neutrino production through the \textit{pp}-channel, but it could also attenuate a significant fraction of the gamma rays that are produced in this process. This could relieve the apparent tension between the diffuse IceCube neutrino flux and the diffuse gamma-ray flux measured by \textit{Fermi}-LAT. We present our source selection criteria and IceCube sensitivities in view of a search for high-energy neutrinos from these so far unexplored objects.\\

\vspace{4mm}
{\bfseries Corresponding authors:}
\speaker{Pablo Correa}$^{1}$, Krijn D.~de Vries$^{1}$, Nick van Eijndhoven$^{1}$\\
{$^{1}$ \itshape Vrije Universiteit Brussel, Pleinlaan 2, BE-1050 Brussels, Belgium}\\

}

\FullConference{36th International Cosmic Ray Conference -ICRC2019-\\
		July 24th - August 1st, 2019\\
		Madison, WI, U.S.A.}

\usepackage{caption}
\usepackage{subcaption}

\usepackage{amsmath}

\usepackage{lineno}

\begin{document}

\section{Introduction}\label{sec:introduction}

\subsection{Gamma-Ray Obscured Neutrino Sources}

Since the first observation of a diffuse high-energy cosmic neutrino flux with the IceCube observatory \cite{Aartsen:2013_Discovery}, various searches have been performed to identify the origin of these cosmic neutrinos. However, apart from the blazar TXS 0506+056 being identified as the first possible source \cite{Aartsen:2018_TXS,Aartsen:2018_TXSmultimessenger}, the origin of high-energy cosmic neutrinos largely remains unknown. Moreover, the contribution of blazars---observed in gamma rays by the \textit{Fermi}-LAT experiment \cite{Atwood:2009_Fermi}---to the diffuse neutrino flux is strongly constrained \cite{Aartsen:2017_BlazarConstraints}.

Because neutrinos and gamma rays are expected to be produced in the same hadronic processes, one could in first instance expect that the diffuse gamma-ray flux observed by \textit{Fermi}-LAT and the diffuse neutrino flux measured with IceCube have a common origin. However, in contrast to the IceCube neutrino flux, the major contribution to the \textit{Fermi}-LAT gamma-ray flux is due to blazars (about 86\%) \cite{Ackermann:2016_BlazarsFermi}. If one subtracts the blazar component from the diffuse gamma-ray flux and estimates the corresponding non-blazar neutrino flux, the latter underestimates the IceCube data \cite{Bechtol:2017_GammaDeficit}. 

Thus, there seems to be a lack of gamma rays observed with \textit{Fermi}-LAT with respect to the IceCube neutrinos. This tension could be resolved if the sources of high-energy cosmic neutrinos are opaque to gamma rays, i.e.~the gamma rays would be attenuated at the source \cite{Murase:2016_HiddenSources}. Such gamma-ray obscured sources of cosmic neutrinos, or hidden sources, are the main motivation for the study presented here.

\subsection{Ultra-Luminous Infrared Galaxies}
\begin{figure}[ht]
	\begin{center}
		\begin{subfigure}[b]{0.5\textwidth}
			\begin{center}
				\includegraphics[width=\textwidth]{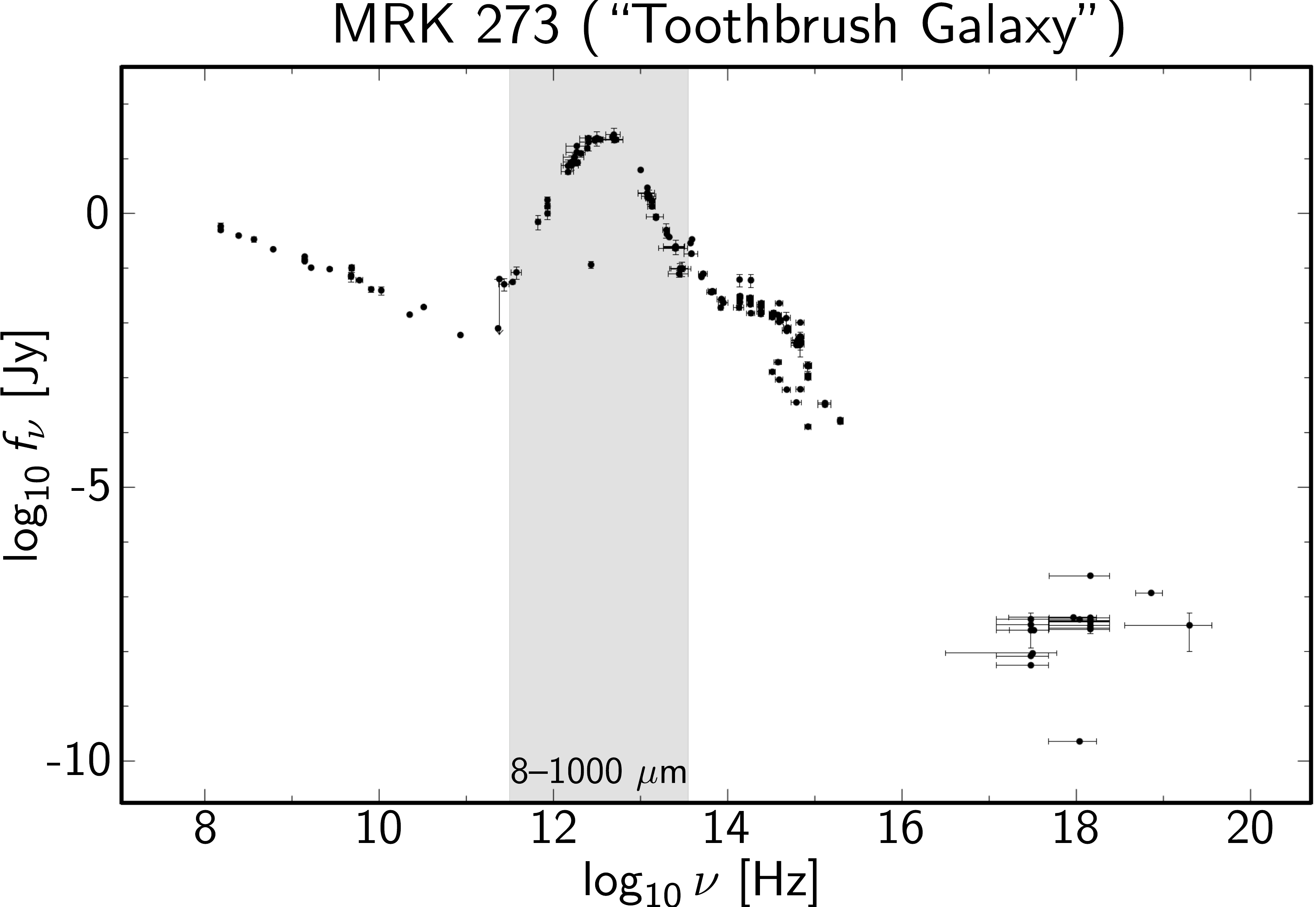}
			\end{center}
		\end{subfigure}
		\hspace{1cm}
		\begin{subfigure}[b]{0.283\textwidth}
			\begin{center}
				\includegraphics[width=\textwidth]{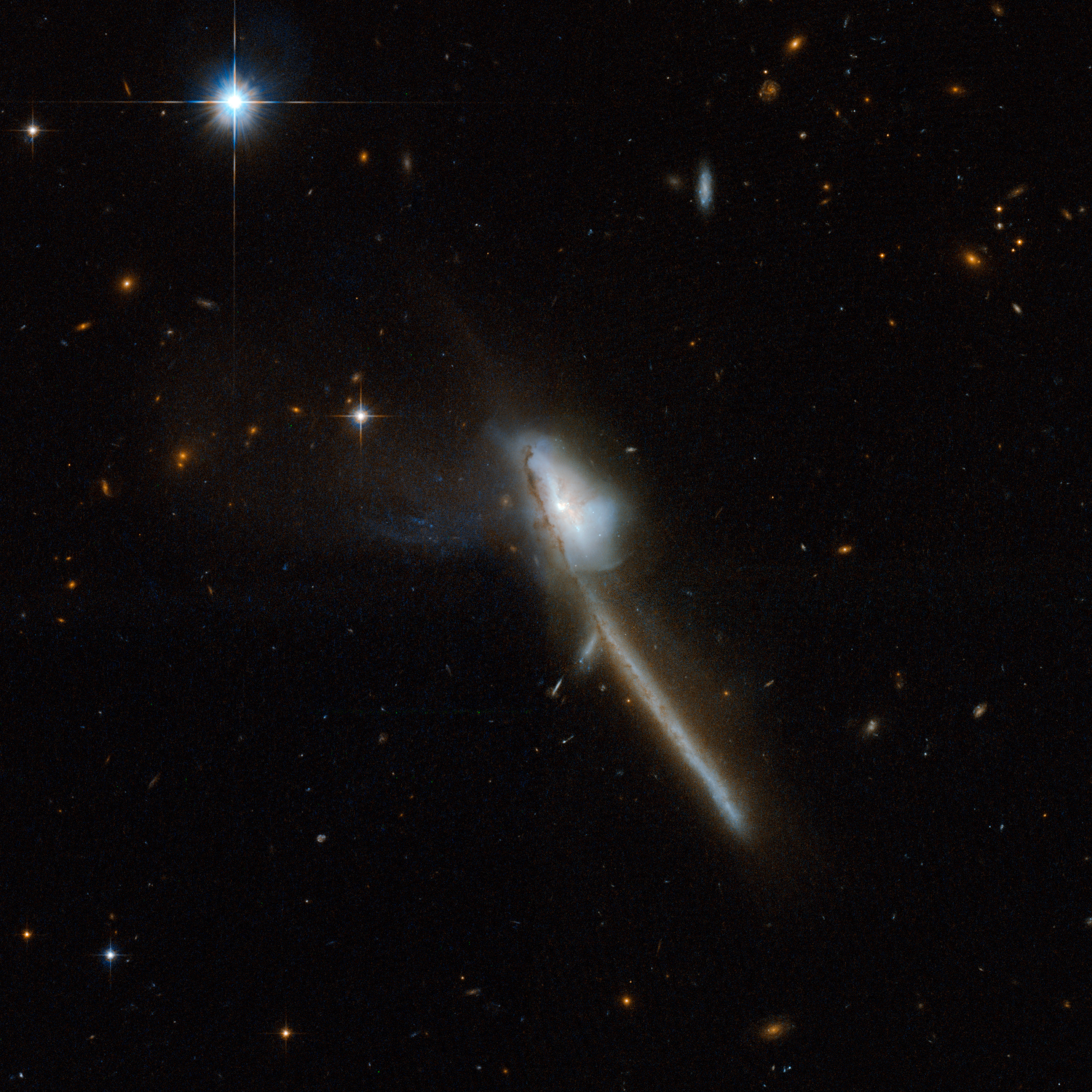}
				\vspace{0.05cm}
			\end{center}
		\end{subfigure}
	\caption{Properties of MRK 273, a typical ULIRG. \textit{Left}: The electromagnetic spectrum of MRK 273, where the frequency is denoted by $\nu$ and the flux at that frequency is denoted by $f_{\nu}$. The shaded area indicates the 8--1000 $\mu$m infrared band, in which the dominant infrared emission is clearly visible. Figure taken and adapted from NED \cite{NED}. \textit{Right}: An optical image of MRK 273, also known as the ``Toothbrush Galaxy'', shows a pair of interacting galaxies, which is another typical feature of ULIRGs. Credit: ESA/Hubble.}
	\label{fig:MRK273}
	\end{center}
\end{figure}

Ultra-Luminous Infrared Galaxies (ULIRGs) have infrared luminosities $L_{\mathrm{IR}} \geq 10^{12}~ L_{\odot} \equiv L_{\odot}^{12}$ in the 8--1000 $\mu$m wavelength range, with $L_{\odot}$ the solar luminosity, making them the most luminous objects in the infrared sky (see \cite{Lonsdale:2006_ULIRGreview} for a review). A typical electromagnetic spectrum of an ULIRG is shown in the left panel of Fig.~\ref{fig:MRK273}, taken from the NASA/IPAC Extragalactic Database\footnote{The NASA/IPAC Extragalactic Database (NED) is operated by the Jet Propulsion Laboratory, California Institute of Technology, under contract with the National Aeronautics and Space Administration.} (NED) \cite{NED}, visualizing the dominant infrared contribution. The strong infrared emission indicates that there are large quantities of dust in these objects, as dust emits thermal infrared radiation when heated.

The main energetic environments in ULIRGs responsible for these extreme infrared luminosities are starburst regions with star formation rates up to $1000~M_{\odot}~\mathrm{yr}^{-1}$ \cite{Lonsdale:2006_ULIRGreview} (the current star formation rate of the Milky Way is about $1~M_{\odot}~\mathrm{yr}^{-1}$ \cite{Robitaille:2010_GalacticSFR}). In addition, there is evidence for the presence of Active Galactic Nuclei (AGN) in ULIRGs, which can contribute significantly to the total infrared luminosity \cite{Nardini:2010_PSCz}. The morphology of ULIRGs further indicates that these objects host extreme environments, since they are typically interacting galaxies \cite{Hung:2014_Morphology}, as exemplified in the right panel of Fig.~\ref{fig:MRK273}, showing an optical image of MRK 273 by ESA/Hubble\footnote{Image taken from \href{https://www.spacetelescope.org/images/heic0810av/}{https://www.spacetelescope.org/images/heic0810av/}.}.

If hadronic acceleration occurs within ULIRGs, both high-energy neutrinos and gamma rays are expected to be produced within these energetic environments, for example in the interactions of high-energy protons with ambient photons through the photohadronic $p\gamma$-channel. However, the gamma rays could be attenuated by the abundant dust present in these objects. Moreover, this dust could act as a target for the accelerated protons, which would lead to an additional production of neutrinos through the hadronic $pp$-channel. Therefore, ULIRGs are ideal gamma-ray obscured source candidates of high-energy cosmic neutrinos.

\section{Selection of Ultra-Luminous Infrared Galaxies}\label{sec:selection}

In order to perform a first IceCube search for high-energy cosmic neutrinos from ULIRGs, a selection of these objects was made from three catalogs based on data from the Infrared Astronomical Satellite (\textit{IRAS}) \cite{Neugebauer:1984_IRAS}. The three catalogs are described below; each of them provides a value for the total infrared luminosity $L_{\mathrm{IR}}$ of the considered objects. The main requirement for this selection was to obtain objects with $L_{\mathrm{IR}} \geq L_{\odot}^{12}$.

\begin{enumerate}
	\item The \textit{IRAS} Revised Bright Galaxy Sample; here called the Bright Source Catalog (BSC) \cite{Sanders:2003_BSC}. This catalog contains the brightest extragalactic sources observed by \textit{IRAS}, i.e.~objects with a 60 $\mu$m infrared flux $f_{60} > 5.84~\mathrm{Jy}$. It provides the most accurate \textit{IRAS}-based values of $L_{\mathrm{IR}}$ for local infrared sources. The ULIRGs in this catalog were identified as those sources with $L_{\mathrm{IR}} \geq L_{\odot}^{12}$, leading to a selection of 18 objects.
	
	\item The \textit{IRAS} Survey of ULIRGs in the Faint Source Catalog (FSC) \cite{Kim:1998_FSC}. The FSC contains extragalactic objects with $f_{60} > 1~\mathrm{Jy}$, from which the ULIRGs were selected for this survey. The ULIRG sample is unbiased, even though it does not cover the whole sky due to the availability of an optical counterpart for redshift measurements. As such, all 118 ULIRGs were selected from this catalog.
	
	\item The ULIRG catalog from the revised \textit{IRAS} Point Source Catalog (PSCz) combined with \textit{Spitzer} observations \cite{Nardini:2010_PSCz}. This catalog was constructed for a study on the presence of AGN in ULIRGs. It was constructed from ULIRGs identified in the PSCz, containing sources with $f_{60} \gtrsim 0.6~\mathrm{Jy}$, which were also observed by the \textit{Spitzer} spectrometer \cite{Werner:2004_Spitzer}. From this unbiased sample, those ULIRGs were selected with an average value $L_{\mathrm{IR}} \geq L_{\odot}^{12}$. Thus, 163 ULIRGs were selected from this catalog.
	
\end{enumerate}

\begin{figure}[ht]
	\begin{center}
		\includegraphics[width=0.6\textwidth]{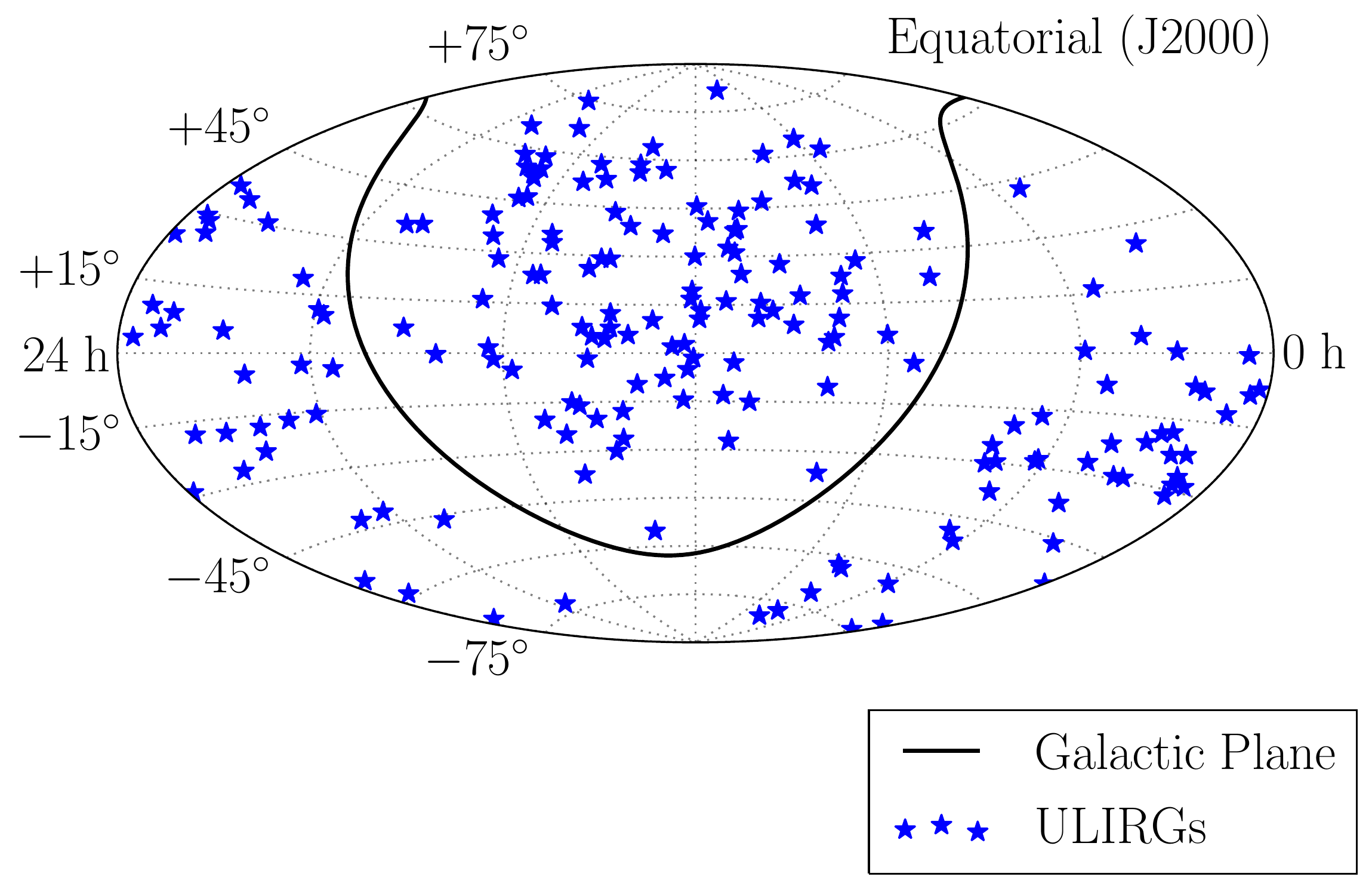}
		\caption{The spatial distribution of the 189 selected ULIRGs. It can be observed that the objects are evenly distributed over the sky, excluding the Galactic Plane.}
		\label{fig:skymap}
	\end{center}
\end{figure}

Because the three catalogs overlap, duplicate objects had to be accounted for, leading to a final selection of 189 ULIRGs which are evenly distributed over the full sky, as shown in Fig.~\ref{fig:skymap}. Since the objects were taken from unbiased catalogs (all objects in catalog 1 can be found in either one of the two other catalogs), the selected ULIRG sample is by construction unbiased.

\begin{figure}[ht]
	\begin{center}
		\begin{subfigure}[b]{0.42\textwidth}
			\begin{center}
			\includegraphics[width=\textwidth]{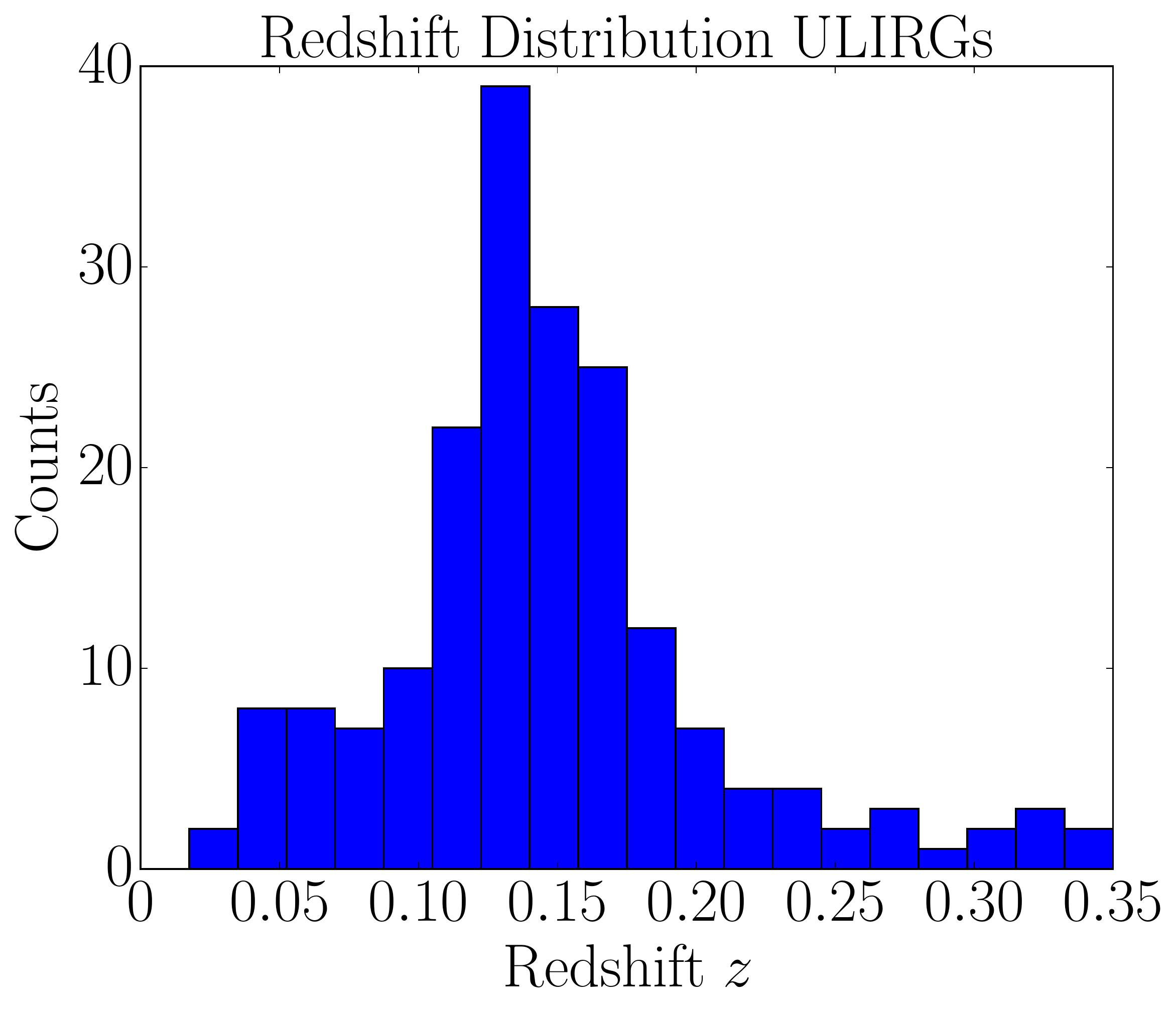}
			\end{center}
		\end{subfigure}
		\hspace{0.05\textwidth}
		\begin{subfigure}[b]{0.42\textwidth}
			\begin{center}
				\includegraphics[width=\textwidth]{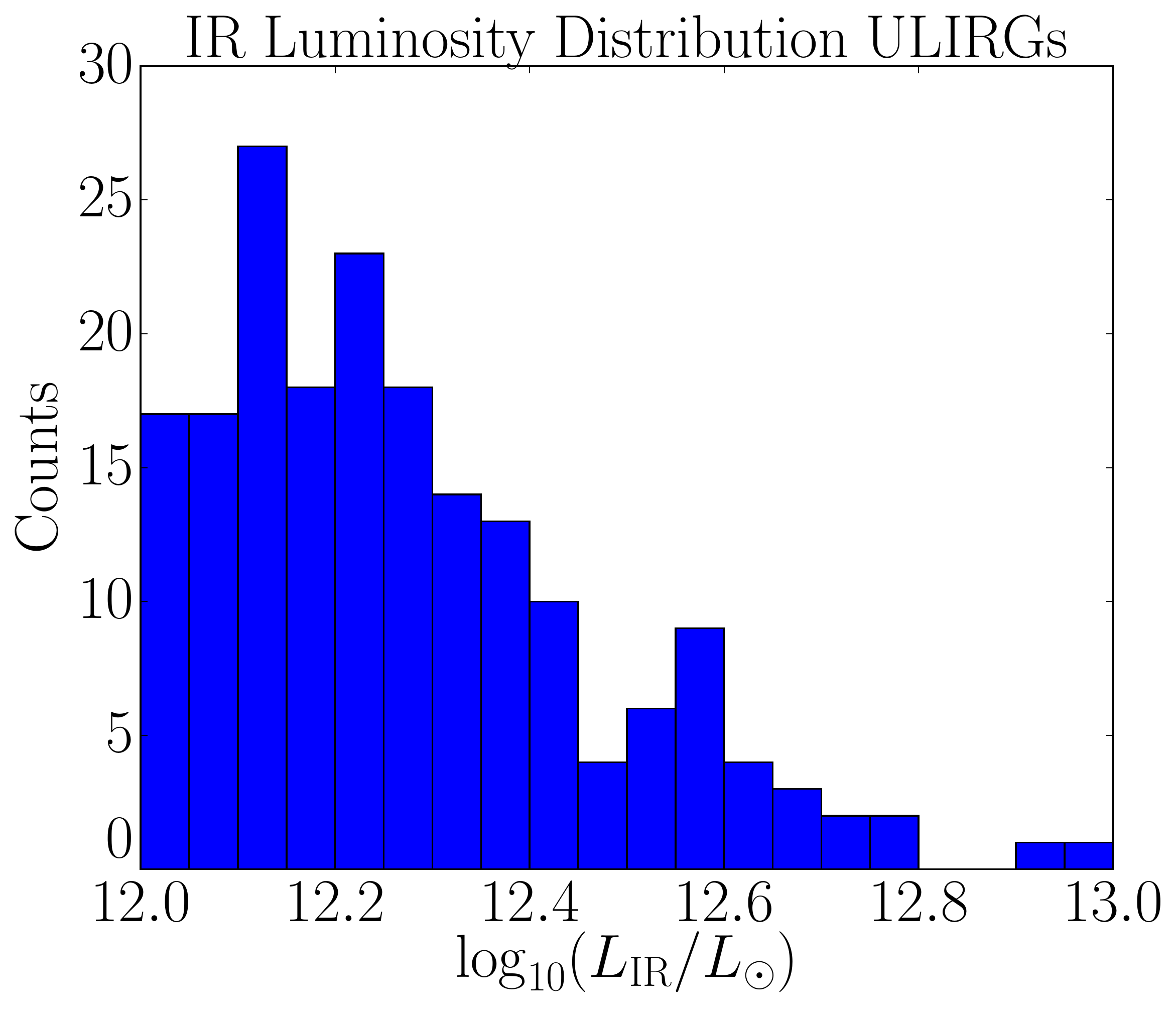}
			\end{center}
		\end{subfigure}
		\caption{Properties of the 189 ULIRGs selected for an IceCube analysis. \textit{Left}: The redshift distribution of the selected ULIRGs, showing that they reside in the relatively nearby Universe. \textit{Right}: The distribution of the total infrared luminosity for the selected objects, which lie within the same order of magnitude.}
		\label{fig:properties}
	\end{center}
\end{figure}

Furthermore, as shown in Fig.~\ref{fig:properties}, the objects are located in the relatively nearby Universe, with redshifts $z<0.35$, and they have total infrared luminosities in the same order of magnitude, $L_{\odot}^{12} \leq L_{\mathrm{IR}} < L_{\odot}^{13}$. The equatorial coordinates and redshift data were taken from the NED, in order to have a homogeneous set of spatial coordinates and redshifts for the ULIRG selection. Additionally, the values of $L_{\mathrm{IR}}$ were taken from catalog 1 if available, otherwise from catalog 2 if available, and finally from catalog 3 if the object was not in the previous catalogs.

\section{IceCube Search for Neutrinos from Ultra-Luminous Infrared Galaxies}

\subsection{Stacking Analysis Method}

The $1~ \mathrm{km}^3$ IceCube neutrino observatory, located at the geographic South Pole, contains 5160 optical modules which detect the optical Cherenkov radiation emitted by secondary charged particles produced in the interactions of neutrinos in the surrounding ice or the nearby bedrock. Using the detected Cherenkov light, the direction, energy and flavor of the neutrino can be reconstructed \cite{Aartsen:2017_Detector}.

The main background in a search for astrophysical neutrinos with IceCube is due to atmospheric muons and atmospheric neutrinos produced in cosmic-ray air showers. The former trigger the detector with a rate of about 2.7 kHz, thus stringent filters are applied to select only those events with a high likelihood of being produced in neutrino interactions. This reduces the background rate to the atmospheric neutrino level, which is about 6.7 mHz. However, a statistical analysis is required to search for an astrophysical signal component in the data, since the atmospheric neutrinos can typically not be distinguished from astrophysical neutrinos on an event-by-event basis.

For the search of high-energy cosmic neutrinos from ULIRGs, a time-integrated unbinned maximum likelihood analysis is performed \cite{Braun:2008_LLH}, where the likelihood takes the form
\begin{equation}
	\mathcal{L} = \prod_{i=1}^{N} \left[ \frac{n_s}{N} S(x_i,\gamma) + \left( 1- \frac{n_s}{N} \right)  B(x_i) \right].
	\label{eq:llh}
\end{equation}
Here, the likelihood is constructed from $N$ events, using their reconstructed parameters $x_i$ which contain both spatial and energy information. In addition, $n_s$ denotes the amount of signal events. As such, both the background Probability Density Function (PDF), $B$, and signal PDF, $S$, contain a spatial and energy component, where the energy component of $S$ depends on the spectral index $\gamma$ of the signal, which is assumed to follow a power-law spectrum.

In this analysis, instead of considering the $M=189$ ULIRGs separately, the corresponding signal PDFs $S_k$ are stacked together in order to enhance the sensitivity of this search. The stacking is performed in the form of a weighed sum,
\begin{equation*}
	S(x_i,\gamma) = \frac{\sum_{k=1}^{M} W_k R_k S_k(x_i,\gamma)}{\sum_{k=1}^{M} W_k R_k},
\end{equation*}
which is the signal PDF that enters the likelihood in Eq.~\ref{eq:llh}. The weights are separated into detector weights $R_k$, which take into account the performance of the detector at a certain ULIRG position on the sky, and theoretical weights $W_k$.

Since the selected ULIRGs have total infrared luminosities within the same order of magnitude, $L_{\odot}^{12} \leq L_{\mathrm{IR}} < L_{\odot}^{13}$, it is assumed that the ULIRGs have a similar neutrino luminosity, i.e~we assume ULIRGs to be neutrino standard candles. Consequently, since the redshifts of the ULIRGs are known, the contribution of each ULIRG to their cumulative neutrino flux at Earth can be accounted for. Therefore, for this analysis it is chosen that $W_k = d_k^{-2}$, where $d_k$ denotes the comoving distance to the $k^{\mathrm{th}}$ ULIRG. The distances were determined from the redshift data using the \textit{Planck} 2015 cosmology measurements (with a cosmological parameter $H_0 = (67.8 \pm 0.9)~ \mathrm{km~ s^{-1}~ Mpc^{-1}} $) \cite{Ade:2016_Planck}.

The likelihood formulation of Eq.~\ref{eq:llh} allows for the construction of a test statistic,
\begin{equation*}
	\mathrm{TS} = 2 \log \left[ \frac{\mathcal{L}(n_s=\hat{n}_s,\gamma=\hat{\gamma})}{\mathcal{L}(n_s=0)} \right],
\end{equation*}
where $\hat{n}_s$ and $\hat{\gamma}$ are the values that maximize the likelihood. Note that it is assumed that the spectral indices are the same for all ULIRGs, so a single value of $\gamma$ is fit in this analysis. This test statistic allows for the distinction of the hypothesis of a signal component in the data from the background-only hypothesis in terms of a p-value. However, the background only test statistic distribution is required in order to make this p-value estimation. This distribution is determined by performing pseudo-experiments. Each pseudo-experiment consists of determining the test statistic after scrambling the data in right ascension---the IceCube detector is symmetric in right ascension for the timescales considered here---in order to probe random background fluctuations.

\subsection{Sensitivities and Discovery Potentials}

In order to test the performance of the ULIRG stacking analysis in the presence of signal, pseudo-experiments were performed with the additional injection of pseudo-signal. The injection is performed according to
\begin{equation}
	\frac{\mathrm{d}N}{\mathrm{d}E} = A \left( \frac{E}{E_0} \right)^{-\gamma},
	\label{eq:injection}
\end{equation}
with $E$ the neutrino energy and $\mathrm{d}N/\mathrm{d}E$ the differential flux at that energy, which is equal to the normalization $A$ at $E_0$. Here, the normalization energy was chosen to be $E_0 = 1~\mathrm{TeV}$, and the spectral indices of the injected spectrum were chosen from the range $\gamma \in [1.5,3.5]$ in steps of 0.5.

The sensitivity of this analysis is defined as the amount of injected signal required such that in 90\% of the pseudo-experiments, one obtains a p-value $p \leq 0.5$. In addition, the discovery potential is defined as the injected signal required such that in 50\% of the pseudo-experiments, a p-value $p \leq 5.73\times10^{-7}$ is obtained, which corresponds to a $5\sigma$ significance.

\begin{figure}[ht]
	\begin{center}
		\begin{subfigure}[b]{0.48\textwidth}
			\begin{center}
				\includegraphics[width=\textwidth]{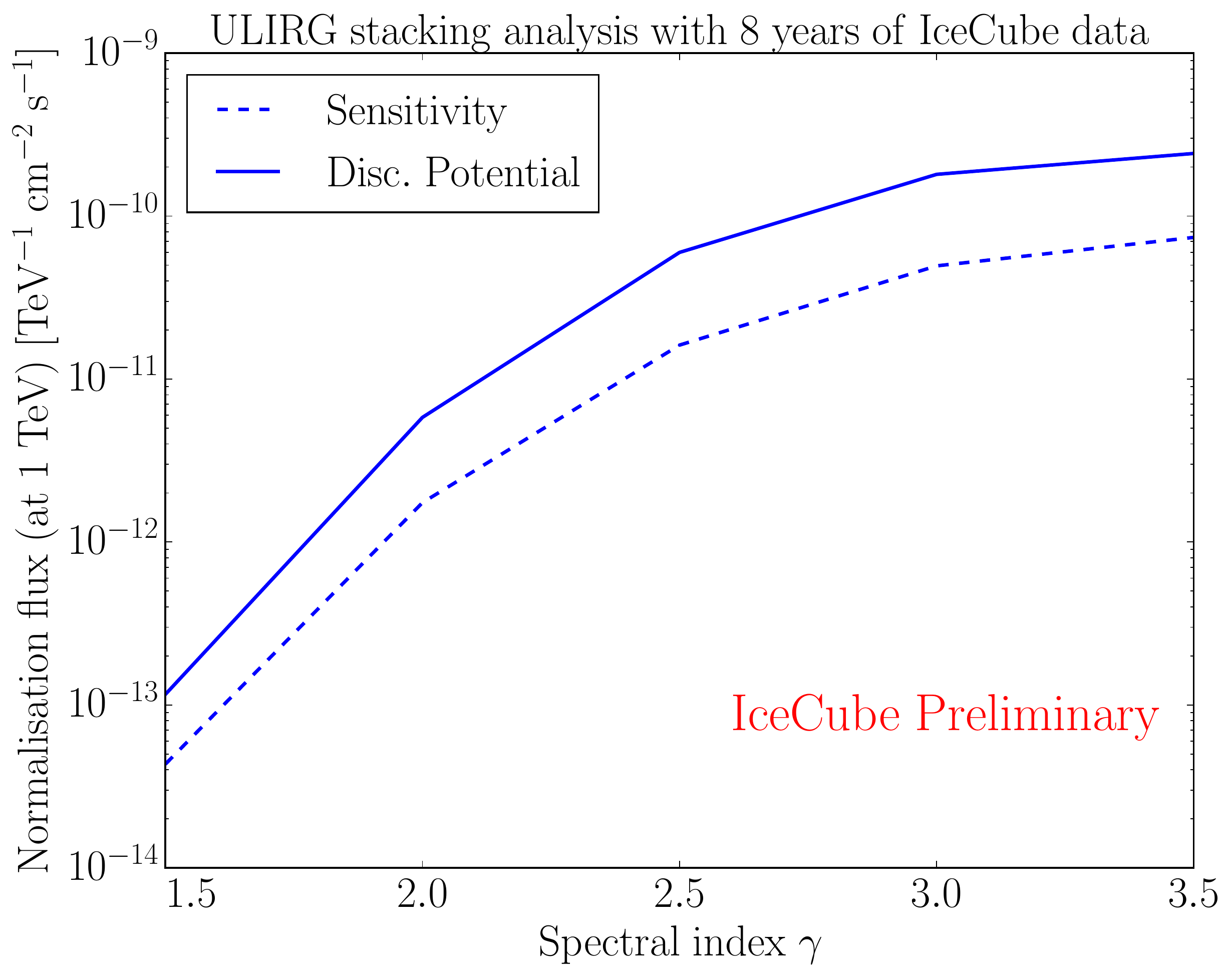}
			\end{center}
		\end{subfigure}
		\hspace{0.02\textwidth}
		\begin{subfigure}[b]{0.48\textwidth}
			\begin{center}
				\includegraphics[width=\textwidth]{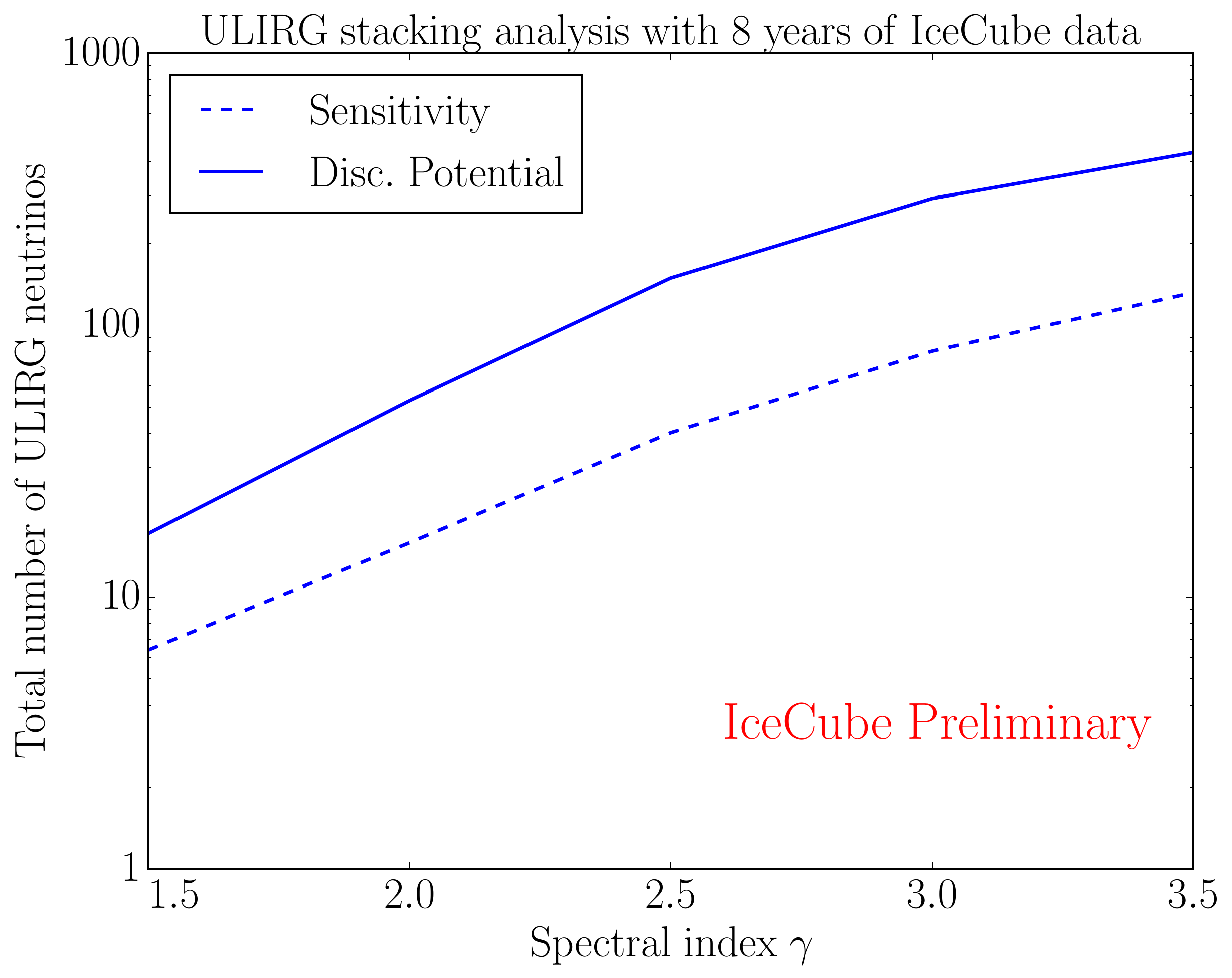}
			\end{center}
		\end{subfigure}
		\caption{Sensitivities and $5\sigma$ discovery potentials for the ULIRG stacking analysis, as a function of the spectral index $\gamma$, for 8 years of IceCube data. \textit{Left}: The sensitivities and discovery potentials in terms of flux normalization. \textit{Right}: The corresponding sensitivities and discovery potentials in terms of the total number of neutrinos from the selected ULIRGs, expected in the energy range from about 10 GeV to 70 PeV.}
		\label{fig:sensitivity}
	\end{center}
\end{figure}

Fig.~\ref{fig:sensitivity} shows the preliminary sensitivities and discovery potentials of the ULIRG stacking analysis as a function of the spectral index $\gamma$, for 8 years of IceCube data. The left panel of Fig.~\ref{fig:sensitivity} shows these quantities in terms of the flux normalization $A$ of Eq.~\ref{eq:injection}, while in the right panel these are shown in terms of the corresponding number $n_{\nu}$ of high-energy neutrinos. This is obtained as
\begin{equation*}
	n_{\nu} = \int \left( \int_{E_1}^{E_2} A_{\mathrm{eff}} \frac{\mathrm{d}N}{\mathrm{d}E} \mathrm{d}E \right) \mathrm{d}t~ \mathrm{d}\Omega,
\end{equation*}
where the differential flux $\mathrm{d}N/\mathrm{d}E$ is convoluted with the effective area of IceCube, $A_{\mathrm{eff}}$, by integrating over the energy range $[E_1 \simeq 10~\mathrm{GeV},E_2 \simeq 70~\mathrm{PeV}]$, the livetime $t$ of the data sample and the solid angle $\Omega$.

It is clear that the analysis is more sensitive for harder spectra, i.e.~lower values of $\gamma$. This is expected, since harder spectra will lead to more neutrinos with higher energy, where the background falls more rapidly given its rather soft spectrum (roughly $E^{-3.7}$). Generally, the conclusion that can be drawn is that the analysis typically requires between 10 and 100 neutrinos in the energy range from about 10 GeV to 70 PeV in terms of sensitivity.

The results of this analysis will have strong implications for the hadronic processes in ULIRGs, in particular for the production of neutrinos through the $pp$-interaction channel, since this is the first IceCube analysis performed on ULIRGs. In the most optimistic scenario, ULIRGs will be identified as sources of high-energy cosmic neutrinos, which might also resolve the tension between the IceCube diffuse neutrino flux and \textit{Fermi}-LAT diffuse gamma-ray flux. Alternatively, if no high-energy neutrinos are found that originate from these objects, stringent upper limits will be set on the cumulative neutrino flux of the 189 ULIRGs. 

\section{Conclusions and Outlook}

ULIRGs not only are the most luminous objects in the infrared sky with total infrared luminosities $L_{\mathrm{IR}} \geq L_{\odot}^{12}$, they are also ideal gamma-ray obscured neutrino source candidates. Such sources could resolve the tension observed between the diffuse gamma-ray flux measured with \textit{Fermi}-LAT and the high-energy neutrino flux detected with IceCube, where there appears to be a lack of gamma rays with respect to the high-energy cosmic neutrinos.

As such, a selection of ULIRGs was made for a search of high-energy neutrinos from these objects with IceCube. The ULIRGs were obtained from three catalogs based on \textit{IRAS} data, leading to a selection of 189 objects evenly distributed over the sky. In addition, the ULIRG sample is unbiased, and with redshifts $z<0.35$ it is restricted to the relatively nearby Universe.

Furthermore, a first sensitivity study was performed on this selection of ULIRGs using 8 years of IceCube data. Depending on the spectral index of the astrophysical neutrinos possibly originating from these objects, the sensitivity of the analysis mostly lies between 10 and 100 neutrinos distributed over the 189 ULIRGs, with energies between about 10 GeV and 70 PeV.

This novel IceCube search for high-energy cosmic neutrinos originating from ULIRGs will allow us to obtain insights in the high-energy environments at these sources. On the one hand, ULIRGs could be identified as sources of high-energy cosmic neutrinos. On the other hand, if no neutrinos are observed from these objects, the very first limits will be set on the high-energy neutrino flux originating from ULIRGs.


\bibliographystyle{ICRC}
\bibliography{references}

\providecommand{\href}[2]{#2}\begingroup\raggedright\begin{thebibliography}{10}

\bibitem{Aartsen:2013_Discovery}
{\bf IceCube} Collaboration, M.~G. Aartsen et~al., {\em Science} {\bf 342}
  (2013) 1242856.

\bibitem{Aartsen:2018_TXS}
{\bf IceCube} Collaboration, M.~G. Aartsen et~al., {\em Science} {\bf 361}
  (2018) 147--151.

\bibitem{Aartsen:2018_TXSmultimessenger}
{\bf IceCube et al.} Collaboration, M.~G. Aartsen et~al., {\em Science} {\bf
  361} (2018) eaat1378.

\bibitem{Atwood:2009_Fermi}
{\bf \textit{Fermi}-LAT} Collaboration, W.~B. Atwood et~al., {\em The
  Astrophysical Journal} {\bf 697} (2009) 1071--1102.

\bibitem{Aartsen:2017_BlazarConstraints}
{\bf IceCube} Collaboration, M.~G. Aartsen et~al., {\em The Astrophysical
  Journal} {\bf 835} (2017) 45.

\bibitem{Ackermann:2016_BlazarsFermi}
{\bf \textit{Fermi}-LAT} Collaboration, M.~Ackermann et~al., {\em Phys. Rev.
  Lett.} {\bf 116} (2016) 151105.

\bibitem{Bechtol:2017_GammaDeficit}
K.~Bechtol et~al., {\em The Astrophysical Journal} {\bf 836} (2017) 47.

\bibitem{Murase:2016_HiddenSources}
K.~Murase, D.~Guetta, and M.~Ahlers, {\em Phys. Rev. Lett.} {\bf 116} (2016)
  071101.

\bibitem{Lonsdale:2006_ULIRGreview}
C.~J. Lonsdale, D.~Farrah, and H.~E. Smith, {\em Ultraluminous Infrared
  Galaxies}, pp.~285--336.
\newblock Springer Berlin Heidelberg, Berlin, Heidelberg, 2006.

\bibitem{NED}
{NASA/IPAC Extragalactic Database, available at
  \href{http://ned.ipac.caltech.edu}{http://ned.ipac.caltech.edu}.}

\bibitem{Robitaille:2010_GalacticSFR}
T.~P. Robitaille and B.~A. Whitney, {\em The Astrophysical Journal} {\bf 710}
  (2010) L11--L15.

\bibitem{Nardini:2010_PSCz}
E.~Nardini et~al., {\em MNRAS} {\bf 405} (2010) 2505--2520.

\bibitem{Hung:2014_Morphology}
C.-L. Hung et~al., {\em The Astrophysical Journal} {\bf 791} (2014) 63.

\bibitem{Neugebauer:1984_IRAS}
G.~{Neugebauer} et~al., {\em The Astrophysical Journal Letters} {\bf 278}
  (1984) L1--L6.

\bibitem{Sanders:2003_BSC}
D.~B. Sanders et~al., {\em The Astronomical Journal} {\bf 126} (2003)
  1607--1664.

\bibitem{Kim:1998_FSC}
D.-C. Kim and D.~B. Sanders, {\em The Astrophysical Journal Supplement Series}
  {\bf 119} (1998) 41--58.

\bibitem{Werner:2004_Spitzer}
M.~W. {Werner} et~al., {\em The Astrophysical Journal Supplement Series} {\bf
  154} (2004) 1--9.

\bibitem{Aartsen:2017_Detector}
{\bf IceCube} Collaboration, M.~G. Aartsen et~al., {\em JINST} {\bf 12} (2017)
  P03012.

\bibitem{Braun:2008_LLH}
J.~Braun et~al., {\em Astroparticle Physics} {\bf 29} (2008) 299 -- 305.

\bibitem{Ade:2016_Planck}
{\bf \textit{Planck}} Collaboration, P.~A.~R. Ade et~al., {\em A\&A} {\bf 594}
  (2016) A13.

\end{thebibliography}\endgroup

\end{document}